\documentstyle[preprint,aps]{revtex}

\begin{document}
\title{{\bf An alternative simple solution of the sextic anharmonic oscillator and
perturbed Coulomb problems }}
\author{Sameer M. Ikhdair\thanks{%
sikhdair@neu.edu.tr} and \ Ramazan Sever\thanks{%
sever@metu.edu.tr}}
\address{$^{\ast }$Department of Physics, \ Near East University, Nicosia, North
Cyprus, Mersin-10, Turkey\\
$^{\dagger }$Department of Physics, Middle East Technical \ University,
06531 Ankara, Turkey.}
\date{\today
}
\maketitle

\begin{abstract}
Utilizing an appropriate ansatz to the wave function, we reproduce the exact
bound-state solutions of the radial Schr\"{o}dinger equation to various
exactly solvable sextic anharmonic oscillator and confining perturbed
Coulomb models in $D$-dimensions. We show that the perturbed Coulomb problem
with eigenvalue $E$ can be transformed to a sextic anharmonic oscillator
problem with eigenvalue $\widehat{E}.$ We also check the explicit relevance
of these two related problems in higher-space dimensions. It is shown that
exact solutions of these potentials exist when their coupling parameters
with $k=D+2\ell $ appearing in the wave equation satisfy certain constraints.


PACS\ number: 03.65.-w; 03.65.Fd; 03.65.Ge.
\end{abstract}


\section{Introduction}

\noindent The solution of the fundamental dynamical equations is an
interesting phenomenon because of its importance in quantum-field theory,
molecular physics, solid-state and statistical physics. To obtain the exact $%
\ell $-state solutions of the Schr\"{o}dinger equation (SE) to various
quantum mechanical problems are possible only for few potentials and hence
approximation methods are used to obtain their solutions. According to the
Schr\"{o}dinger formulation of quantum mechanics, a total wave function
provides implicitly all relevant information about the behaviour of a
physical system. Hence if it is exactly solvable for a given potential, the
wave function can describe such a system completely. Until now, many efforts
have been made to solve the stationary SE with a sextic anharmonic
oscillator and perturbed Coulomb potentials in one to three dimensions
through the Hill determinant matrix method [1-4]. The study of the SE with
these potentials provides us with insight into the physical problem under
consideration. Further, the study of SE with some of these potentials in the
arbitrary dimensions $D$ is presented [3].

The purpose of this paper is to carry out the analytical solutions of the $D$%
-dimensional radial SE with exactly-solvable Coulomb plus linear plus
harmonic (CLH) $V_{1}(r)=-a/r+br+cr^{2}$ and sextic anharmonic oscillator
(AHO) $V_{2}(r)=\mu r^{2}+\lambda r^{4}+\eta r^{6}$ potentials through an
appropriate ansatz to the wave function. In cases of exactly solvable
models, the wave function can be expressed as a finite power series
polynomial multiplied by an appropriate reference function (usually, the
asymptotic form) to reproduce the exact solutions. The analytical solution
of the Schr\"{o}dinger equation for the energy levels with a class of
confining potentials of type $V_{1}$ have been studied by Datta and
Mukherjee [5] by using Kato-Rellich perturbation theory for linear
operations. It is well known that this confinement potential has been used
for calculation of $q\overline{q}$ bound-state masses [6]. Killingbeck [7]
has calculated the energy eigenvalues of the confinement potential by using
hypervirial relations. Exact solutions [8] of potentials of type $V_{1}$ and
$V_{2}$ are obtained by a number of authors in three-dimensional space when
the coupling parameters satisfy certain relations. \ Chaudhuri and Mondal
[3] studied the $D$-dimensional sextic AHO and CLH problems within the
framework of supersymmetric quantum mechanics (SUSYQM) and Hill determinant
method [4,9-11]. Chaudhuri and Mondal have shown that SUSYQM yields exact
solutions for a single state only for the quasi-exactly-solvable potentials
of type $V_{1}$ and $V_{2}$ in $D$-dimensions with some constraints on the
coupling constants [3]. They have also obtained numerical results throughout
the Hill determinant method. The ideas of supersymmetric quantum mechanics
have been used for the study of atomic systems [11], the evaluation of the
eigenvalues of a bistable potential [12], the improvement of the large-N
expansion [13], the analysis of all known shape invariant potentials
[10,14], and the development of a more accurate WKB approximation [14].
Tymczak et al [4] devised a highly accurate quantization procedures for the
inner product representation both in configuration and momentum space for
various AHO potentials in one and two dimensions. Additionally, Dobrovolska
and Tutik [15] studied the bound-state problem within the framework of the
SE through the logarithmic perturbation theory. Recently, they also extended
the formalism to the bound-state problem for spherical oscillator of type $%
V_{2}$ with its subsequent application to the doubly anharmonic oscillator
[16]. Furthermore, a simple formalism [17] based on a suitable choice of the
wave function ansatz has been proposed for reproducing exact bound-state
energy eigenvalues and eigenfunctions for exactly solvable model within the
framework of the SE. Very recently, this simple approach has also been
applied [18] with remarkable success, to various molecular quantum
mechanical problems in $D$-dimensions [19,20].

The object of this paper is to extend the above simple approach to reproduce
bound-state exact solutions for potentials of type $V_{1}$ and $V_{2}$ in $D$%
-dimensions with some constraints on the coupling constants. For a certain
choice of parameters the method provides exactly-solvable potentials for
the\ $D$-dimensional sextic AHO and CLH problems. We then compare our
results with the exact ones obtained from the transformation of LHO into
sextic AHO problem.

This paper is organized as follows. In Section \ref{TDD}, we solve
analytically the $D$-dimensional radial Schr\"{o}dinger equation for the
sextic AHO and CLH problems by a suitable choice of a wave function ansatz
to each exactly-solvable problem. On the other hand, the exact energy
eigenvalues of sextic AHO are obtained by transforming radial wave SE in $D$%
-dimensions with angular momentum $\ell $ to another problem in $(2D-4)$%
-dimensions with angular momentum $2\ell +1.$ The results and conclusion
will be given in Section \ref{CR}.

\section{The $D$-Dimensional Radial Schr\"{o}dinger Equation}

\label{TDD}In the $D$-dimensional Hilbert space, the reduced radial wave
Schr\"{o}dinger equation (with $\hbar =m=1$ units) for a spherically
symmetric potential $V(r)$ takes the form [21]
\begin{equation}
\left[ \frac{d^{2}}{dr^{2}}+\frac{\left( D-1\right) }{r}\frac{d}{dr}-\frac{%
\ell (\ell +D-2)}{r^{2}}+2\left( E-V(r)\right) \right] \psi (r)=0,
\end{equation}
where the interaction potential is chosen to be of type $V_{1}$ or $V_{2}$
and $E$ stands for its eigenvalues. Further, equation (1) can be simply
transformed to the form [21]

\begin{equation}
\left\{ \frac{d^{2}}{dr^{2}}-\frac{\left[ \left( k-1\right) \left(
k-3\right) \right] }{4r^{2}}+2\left( E-V(r)\right) \right\} R(r)=0,
\end{equation}
where $R(r),$ the reduced radial wave function, is defined by

\begin{equation}
R(r)=r^{(D-1)/2}\psi (r),
\end{equation}
and

\begin{equation}
k=D+2\ell ,
\end{equation}
which is a parameter depends on a linear combination of the spatial
dimensions $D$ and the angular momentum quantum number $\ell $ [21]$.$

We substitute $r=\gamma \rho ^{2}/2$ and $\psi =\chi (\rho )/\rho $ to
transform Eq.(1) into another Schr\"{o}dinger-type equation in ($D^{\prime
}=2D-4)$-dimensional space with angular momentum $L=2\ell +1,$

\begin{equation}
\left[ \frac{d^{2}}{d\rho ^{2}}+\frac{\left( D^{\prime }-1\right) }{\rho }%
\frac{d}{d\rho }-\frac{L(L+D^{\prime }-2)}{\rho ^{2}}+2\left( \widehat{E}-%
\widehat{V}(\rho )\right) \right] \chi (\rho )=0,
\end{equation}
where

\begin{equation}
\widehat{E}=\gamma ^{2}\rho ^{2}E,\text{ }\widehat{V}(\rho )=\gamma ^{2}\rho
^{2}V(\gamma \rho ^{2}/2),\text{ }\gamma =1/(-E)^{1/2}.
\end{equation}
It is seen that through this transformation, the $D$-dimensional radial wave
Schr\"{o}dinger equation (1) with angular momentum $\ell $ can be
transformed to a ($D^{\prime }=2D-4)$-dimensional problem with new angular
momentum $L=2\ell +1.$ In particular, under this transformation, CLH problem
of type $V_{1}$ with eigenvalue $E$ can be transformed to a sextic AHO
problem of type $V_{2}$ with eigenvalue $\widehat{E}$ as
\begin{equation}
\widehat{V}(\rho )=\mu \rho ^{2}+\lambda \rho ^{4}+\eta \rho ^{6},
\end{equation}
with coupling constants given by

\begin{equation}
\mu =1,\text{ }\lambda =\frac{b}{2(-E)^{3/2}},\text{ }\eta =\frac{c}{%
4(-E)^{2}},
\end{equation}
and eigenvalue

\begin{equation}
\widehat{E}=\frac{2a}{(-E)^{1/2}}.
\end{equation}

\subsection{Confining perturbed Coulomb problem}

We attempt to solve the wave equation (2) of reduced radial wave $R(r)$ in
the $D$-dimensions for a spherically symmetric potential of confining CLH
form [3,22]:
\begin{equation}
V(r)=-\frac{a}{r}+br+cr^{2},\text{ }c>0.
\end{equation}
In particular, for $c=0$ and $b>0,$ such a potential reduces to the well
known quarkonium Cornell potential (cf. Refs. [21] and references therein).
Apart from its releavance in heavy quarkonium spectroscopy (cf [21] and
references therein), this class of potentials with $c=0$ has important
applications in atomic physics. For exactly solvable problems such as CLH,
the representation of the radial portion of wave function ansatz, containing
an appropriate reference function (usually, asymptotic form), is
\begin{equation}
R(r)=\exp \left[ p\left( \alpha ,\beta ,r\right) \right]
\sum_{n=0}a_{n}r^{2n+(k-1)/2},
\end{equation}
provided that the power of the reference function has the following
selection:

\begin{equation}
p\left( \alpha ,\beta ,r\right) =\frac{1}{2}\alpha r^{2}+\beta r,
\end{equation}
should fall faster than the asymptotic form of the wave function. Upon
substituting Eq. (11) into Eq. (2) and equating the coefficients of $%
r^{2n+(k-1)/2}$ to zero, we readily arrive at the following relation

\begin{equation}
A_{n}a_{n}+B_{n+1}a_{n+1}+C_{n+2}a_{n+2}=0
\end{equation}
where

\begin{equation}
A_{n}=2E+\beta ^{2}+\alpha \left( 4n+k\right) ,
\end{equation}

\begin{equation}
B_{n}=2a+\beta \left( 4n+k-1\right) ,
\end{equation}
\begin{equation}
C_{n}=4n^{2}+2n(k-2),
\end{equation}
and the value of the parameters for $p\left( \alpha ,\beta ,r\right) $ can
be evaluated as

\begin{equation}
\alpha =\pm \sqrt{2c},\text{ }\beta =\pm \frac{b}{\sqrt{2c}}.
\end{equation}
To obtain a well-behaved solution at the origin and infinity, it is more
convenient to take $\alpha =-\sqrt{2c}$ and $\beta =-\frac{b}{\sqrt{2c}}$
which ensure that wave function ansatz representation in (11), be finite for
all $r$ and convergent at $\infty .$ Further, for a given $p,$ if $a_{p}\neq
0,$ but $a_{p+1}=a_{p+2}=a_{p+3}=\cdots =0,$ we then obtain $A_{p}=0$ from
Eq. (14), i.e.,

\begin{equation}
E_{p}^{D}=-\frac{b^{2}}{4c}+\sqrt{\frac{c}{2}}\left( 4p+2\ell +D\right) ,%
\text{ }p=0,1,2,\cdots .
\end{equation}
Carrying through a parallel analysis to Refs [17,18], $A_{n},$ $B_{n}$ and $%
C_{n}$ must satisfy the determinant relation for a nontrivial solution

\begin{equation}
Det\left|
\begin{array}{cccccc}
B_{0} & C_{1} & \cdots & \cdots & \cdots & 0 \\
A_{0} & B_{1} & C_{2} & \cdots & \cdots & 0 \\
\vdots & \vdots & \vdots & \ddots & \vdots & \vdots \\
0 & 0 & 0 & 0 & A_{p-1} & B_{p}
\end{array}
\right| =0.
\end{equation}
To utilize this method showing the simplicity of this approach, we pursue
determinant analysis to present the exact solution for $p=0,1$ as follows.

Case (1): when $p=0,$ we get from Eq. (18), the exact bound-state solution
of the CLH problem in $D$-dimensions. So that the eigenvalue (ground state)
is given by [17,18]

\begin{equation}
E_{0}^{D}=-\frac{b^{2}}{4c}+\sqrt{\frac{c}{2}}\left( 2\ell +D\right) .
\end{equation}
Further, it is shown from Eq. (19) that $B_{0}=0$, which leads to the
following constraint on the coupling parameters as

\begin{equation}
b(k-1)=2a\sqrt{2c}.
\end{equation}
which consequently, from Eqs. (20) and (21), the perturbed Coulomb potential
admits an exact ground eigen energy$:$

\begin{equation}
E_{0}^{k}=\frac{1}{2}\left[ \frac{b(k-1)^{2}}{2a}+\frac{b(k-1)}{2a}-\frac{%
4a^{2}}{(k-1)^{2}}\right] ,
\end{equation}
which is consistent with Refs [3,22]. \ Further, the corresponding wave
function (unormalized):

\begin{equation}
\psi _{0}^{(k)}(r)=a_{0}r^{\ell }\exp \left[ -\frac{2ar}{(k-1)}-\frac{%
b(k-1)r^{2}}{4a}\right] .
\end{equation}
Case (2): When $p=1,$ the exact energy spectrum becomes

\begin{equation}
E_{1}^{k}=\frac{1}{2}\left[ \frac{b(k+1)^{2}}{2a}+\frac{b(k+1)}{2a}-\frac{%
4a^{2}}{(k+1)^{2}}\right] ,
\end{equation}
and the corresponding wave function (unnormalized) can be readily found as:
\begin{equation}
\psi _{1}^{(k)}(r)=\left( a_{0}+a_{1}r\right) r^{\ell }\exp \left[ -\frac{2ar%
}{(k+1)}-\frac{b(k+1)r^{2}}{4a}\right] ,
\end{equation}
where $a_{0}$ and $a_{1}$ are normalization constants. The relation between
them can be determined through the relation $B_{0}a_{0}+C_{1}a_{1}=0$ to be

\begin{equation}
a_{1}=\left[ \frac{b}{\sqrt{2c}}-\frac{2a}{(k-1)}\right] a_{0}.
\end{equation}
Furthermore, we have the following recurrence relation from Eq. (19), that
is, $B_{0}B_{1}=A_{0}C_{1},$ which consequently provides the following
constraint on the coupling constants of potential:

\begin{equation}
4a^{2}-\frac{b^{2}}{2c}(k-1)(k+1)-\frac{4ab}{\sqrt{2c}}k=\frac{2b}{a}k(k-1).
\end{equation}
Following this approach, we can further generate a class of exact solutions
through setting $p=0,1,2,\cdots ,$ etc. Generally speaking, if $a_{p}\neq 0,$
$a_{p+1}=a_{p+2}=\cdots =0,$ from which we can obtain the energy spectra \
(cf. determinant (19)). For the generalization, one needs to use the shape
invariance property and the relation between supersymmetric partners [22,23]
to find the general solution of energy eigenvalues as
\begin{equation}
E_{n}^{k}=\frac{1}{2}\left[ \frac{b(2n+k-1)^{2}}{2a}+\frac{b(2n+k-1)}{2a}-%
\frac{4a^{2}}{(2n+k-1)^{2}}\right] ,
\end{equation}
with the corresponding wave functions

\begin{equation}
\psi ^{(p)}(r)=\left( a_{0}+a_{1}r+\cdots +a_{p}r^{p}\right) r^{\ell }\exp
\left[ -\frac{2ar}{(2n+k-1)}-\frac{b(2n+k-1)r^{2}}{4a}\right] ,
\end{equation}
where $a_{i}$ $(i=0,1,2,\cdots ,p)$ are normalization constants.

Finally, considering the following exactly solvable potentials. In
particular: (i) Harmonic oscillator: when $a=b=0$ and $c=\frac{1}{2}\omega
^{2}$ are inserted to Eq. (10), giving $\alpha =-\omega $ and $\beta =0.$
Thus, we can readily obtain the energy eigenvalues through using Eqs.
(14)-(16) as [16,18,24]

\begin{equation}
E_{n\ell }=\frac{\omega }{2}\left( 4n+D+2\ell \right) ,\text{ }n,\ell
=0,1,2,\cdots ,
\end{equation}
and the corresponding radial wave function becomes

\begin{equation}
\psi ^{(n)}(r)=\left( a_{0}+a_{1}r^{2}+\cdots +a_{n}r^{2n}\right) r^{\ell
}\exp \left[ -\frac{\omega }{2}r^{2}\right] .
\end{equation}
where $a_{i}$ with $i=0,1,2,\cdots ,n$ are normalization constants.

(ii) Coulomb problem: when $b=c=0,$ and $a=Z,$ implies that $\alpha =0$ and $%
\beta =-\frac{2Z}{(2n+D+2\ell -1)},$ from which we readily obtain the exact
eigenvalues as [25]

\begin{equation}
E_{n\ell }=-\frac{2Z^{2}}{\left( 2n+D+2\ell -1\right) ^{2}},\text{ }n,\ell
=0,1,2,\cdots ,
\end{equation}
and radial wave function becomes

\begin{equation}
\psi ^{(n)}(r)=\left( a_{0}+a_{1}r^{2}+\cdots +a_{n}r^{2n}\right) r^{\ell
}\exp \left[ -\frac{2Z}{(2n+D+2\ell -1)}r\right] ,\text{ }
\end{equation}
where $a_{i}$ with $i=0,1,2,\cdots ,n$ are normalization constants

\subsection{The sextic anharmonic oscillator problem}

This version of sextic AHO potential
\begin{equation}
V(r)=\mu r^{2}+\lambda r^{4}+\eta r^{6},\text{ }\eta >0,
\end{equation}
has been studied in the $D$ dimensions through Hill determinant method [3].
We want to solve the radial SE, \ Eq. (2), with Eq. (34) by selecting the
following representation of ansatz to the radial portion of wave function

\begin{equation}
R(r)=\exp \left[ p\left( \alpha ,\beta ,r\right) \right]
\sum_{n=0}a_{n}r^{2n+(k-1)/2},
\end{equation}
provided that the power of the reference function has the following
selection:

\begin{equation}
p\left( \alpha ,\beta ,r\right) =\frac{1}{2}\alpha r^{2}+\frac{1}{4}\beta
r^{4}.
\end{equation}
Implementing the present method on the representation (35), and taking the
coefficients of $r^{2n+(k+1)/2}$ to zero, we obtain the relation

\begin{equation}
A_{n}a_{n}+B_{n+1}a_{n+1}+C_{n+2}a_{n+2}=0,
\end{equation}
where

\begin{equation}
A_{n}=\alpha ^{2}+\beta \left( 4n+k+2\right) -2\mu ,
\end{equation}

\begin{equation}
B_{n}=2E+\alpha \left( 4n+k\right) ,
\end{equation}
\begin{equation}
C_{n}=2n(2n+k-2),
\end{equation}
and the value of the parameters for $p\left( \alpha ,\beta ,r\right) $ can
be evaluated as

\begin{equation}
\beta =\pm \sqrt{2\eta },\text{ }\alpha =\pm \frac{\lambda }{\sqrt{2\eta }}.
\end{equation}
To obtain a well-behaved solution at the origin and infinity, we must set $%
\beta =-\sqrt{2\eta }$ and $\alpha =-\frac{\lambda }{\sqrt{2\eta }}$ which
ensures that wave function ansatz, Eq. (35), be finite for all $r$. Further,
for a given $p,$ if $a_{p}\neq 0,$ but $a_{p+1}=a_{p+2}=a_{p+3}=\cdots =0,$
we then obtain $A_{p}=0$ from Eq. (38), which leads to the following
constraint on the coupling parameters of sextic AHO problem as
\begin{equation}
2\mu +\sqrt{2\eta }(4n+k+2)-\frac{\lambda ^{2}}{2\eta }=0.
\end{equation}
Case (1): \ when $p=0,$ it is shown from Eq. (19) that $B_{0}=0$, which
leads to the following energy eigenvalue (ground state):

\begin{equation}
E_{0}^{k}=\frac{\lambda k}{2\sqrt{2\eta }},
\end{equation}
and the corresponding wave function (unormalized):

\begin{equation}
\psi _{0}(r)=a_{0}r^{\ell }\exp \left[ -\frac{\lambda r^{2}+\eta r^{4}}{%
\sqrt{2\eta }}\right] ,
\end{equation}
Case (2):\ when $p=1,$ it is shown from Eq. (19) that $%
B_{0}B_{1}=A_{0}C_{1}, $ which leads to the constraints on the coupling
parameters of the potential as

\begin{equation}
E_{1}^{k}=\frac{\lambda }{2\sqrt{2\eta }}(k+2)+\sqrt{\frac{\lambda ^{2}}{%
4\eta }(k+2)-\frac{k}{2}\left[ \sqrt{2\eta }(k+2)+2\mu \right] }.
\end{equation}
In particular, an important version of the above sextic AHO is the harmonic
oscillator problem
\begin{equation}
V(r)=\mu r^{2},\text{ }\mu >0.
\end{equation}
Selecting the following representation of the wave function ansatz::

\begin{equation}
R(r)=\exp \left[ p\left( \alpha ,r\right) \right]
\sum_{n=0}a_{n}r^{2n+(k-1)/2},
\end{equation}
with

\begin{equation}
p\left( \alpha ,r\right) =\frac{1}{2}\alpha r^{2},
\end{equation}
and itirating the previous steps, one gets:

\begin{equation}
A_{n}=\alpha ^{2}-2\mu ,
\end{equation}

\begin{equation}
B_{n}=2E+\alpha \left( 4n+k\right) ,
\end{equation}
\begin{equation}
C_{n}=2n(2n+k-2),
\end{equation}
which leads to choosing $\alpha =-\sqrt{2\mu }.$

Case (1): when $p=0,$ it is shown from Eq. (19) that $B_{0}=0$, which leads
to the following energy eigenvalue (ground state):

\begin{equation}
E_{0}^{k}=\frac{\sqrt{2\mu }k}{2},
\end{equation}
and the corresponding wave function (unormalized):

\begin{equation}
\psi _{0}^{(k)}(r)=a_{0}r^{\ell }\exp \left[ -\frac{\sqrt{2\mu }}{2}r^{2}%
\right] ,
\end{equation}
Case (2): when $p=1,$ it is shown from Eq. (19) that $B_{0}B_{1}=A_{0}C_{1},$
which leads to the restriction on $k$ and the parameters of the potential as

\begin{equation}
E_{1}^{k}=\frac{\sqrt{2\mu }}{2}(k+2)+\sqrt{2\mu }.
\end{equation}
Generally speaking, the energy eigenvalues are

\begin{equation}
E_{n}^{k}=\frac{\sqrt{2\mu }}{2}(4n+k),
\end{equation}
and the corresponding wave function can be read

\begin{equation}
\psi _{n}(r)=\left( a_{0}+a_{1}r+\cdots +a_{p}r^{2n}\right) r^{\ell }\exp
\left[ -\frac{\sqrt{2\mu }}{2}r^{2}\right] ,
\end{equation}
where $a_{i}$ $(i=0,1,2,\cdots ,n)$ are normalization constants.

\section{Concluding Remarks}

\label{CR}We applied the wavefunction ansatz method to the confining CLH and
sextic AHO interactions. Table 1 shows the calculated energies of the CLH
type potential together with those obtained by the SUSYQM and Hill
determinant method for high values of parameters in three- and
four-dimensions. We know from (8) and (9) that the CLH problem with exactly
solvable potentials $V_{1}^{(1),(2),(3)}(r)$ in four-dimensions can be
transformed to the sexrtic AHO problem (7) in four-dimensions with the exact
eigenvalues given through (9).

We compute the energy eigenvalues of the following conjugate AHO potentials:
$\widehat{V}_{1}^{(4)}(r)=r^{2}+\left[ 1/2(7.625)^{3/2}\right] r^{4}$ $+%
\left[ 1/4(32)(7.625)^{2}\right] r^{6},$ $\widehat{V}_{1}^{(5)}(r)=r^{2}+%
\left[ 1/2(7.375)^{3/2}\right] r^{4}$ $+\left[ 1/4(32)(7.375)^{2}\right]
r^{6}$ and $\widehat{V}_{1}^{(6)}(r)=r^{2}+\left[ 1/2(7.125)^{3/2}\right]
r^{4}$ $+\left[ 1/4(32)(7.125)^{2}\right] r^{6}$ in two-dimensions by the
present simple method and compute our results in Table 2 with the exact
values given by (9). A class of AHO may be constructed from (43) that admits
exact solutions. We also compute the energy eigenvalues of the following
conjugate AHO potentials: $\widehat{V}_{1}^{(4)}(r)=r^{2}+\left[
1/2(7.5)^{3/2}\right] r^{4}$ $+\left[ 1/4(32)(7.5)^{2}\right] r^{6},$ $%
\widehat{V}_{1}^{(5)}(r)=r^{2}+\left[ 1/2(7.25)^{3/2}\right] r^{4}$ $+\left[
1/4(32)(7.25)^{2}\right] r^{6}$ and $\widehat{V}_{1}^{(6)}(r)=r^{2}+\left[
1/2(7.0)^{3/2}\right] r^{4}$ $+\left[ 1/4(32)(7.0)^{2}\right] r^{6}$ in
four-dimensions by the method and compute our results in Table 3 with the
exact values given by (9).  These eigenvalues are checked by other methods.

This method yields exact solutions for a single state only for a potential
of type $V_{2}$ and many states of type $V_{2}$ with some constraints on the
coupling parameters. Our method is applicable to many general CLH or AHO and
produces excellent results for the low-lying states. It gives the exact
solutions of the Coulo\i mb and the harmonic oscillator in $D$-dimensions.
The CLH and sextic AHO in $D$-dimensions are related through Eq.(9) and are
verified in Table 2 by this method.

A class of conjugate AHO having exact eigenvalues may be constructed from
the transformation of CLH potential. It is found that the eigenvalues of
central potential $V(r)$ are identical for $D=2$,$\ell =2,$ $D=4,\ell =1$
and $D=6,\ell =0$ states. This is because $k=D+2\ell $ remains unaltered for
these states.

\acknowledgments S.M. Ikhdair is grateful to his family for love and
assistance.

\bigskip

\bigskip \baselineskip= 2\baselineskip
\bigskip

\begin{table}[tbp]
\caption{The lowest energy eigenvalues of the CLH potential for $\ell =0,1,2$
in three- and four-dimensions.}
\begin{tabular}{lllllll}
$a$\tablenotemark[1]\tablenotetext[1]{The parameter values here are taken
from [3].} & $b$ & $c$ & $\ell $ & $D$ & Present method & SUSYQM [3] \\
\tableline$4$ & $1$ & $\frac{1}{32}$ & $0$ & $3$ & $-7.625$ & $-7.625$ \\
$8$ & $1$ & $\frac{1}{32}$ & $1$ & $3$ & $-7.375$ & $-7.375$ \\
$12$ & $1$ & $\frac{1}{32}$ & $2$ & $3$ & $-7.125$ & $-7.125$ \\
$6$ & $1$ & $\frac{1}{32}$ & $0$ & $4$ & $-7.500$ & $-7.500$ \\
$10$ & $1$ & $\frac{1}{32}$ & $1$ & $4$ & $-7.250$ & $-7.250$ \\
14 & $1$ & $\frac{1}{32}$ & $2$ & $4$ & $-7.000$ & $-7.000$%
\end{tabular}
\end{table}

\begin{table}[tbp]
\caption{The eigenvalues of the conjugate sextic anharmonic oscillators in
two-dimensions.are compared with the exact values.}
\begin{tabular}{llll}
Conjugate sextic AHO\tablenotemark[1]\tablenotetext[1]{The parameter values
in constructing the sextic AHO are taken from Table 1.} & $\ell $ & Present
work & Exact Value, Eq.(9) \\
\tableline$\widehat{V}_{1}^{(4)}(r)$ & $1$ & 2$.8971438733606$ & 2$%
.8971438733606$ \\
$\widehat{V}_{1}^{(5)}(r)$ & $3$ & 5$.8916775545493$ & 5$.8916775545493$ \\
$\widehat{V}_{1}^{(6)}(r)$ & $5$ & 8$.9912237911843$ & 8$.9912237911846$%
\end{tabular}
\end{table}

\begin{table}[tbp]
\caption{The eigenvalues of the conjugate sextic anharmonic oscillators in
four-dimensions.are compared with the other works and the exact values.}
\begin{tabular}{lllll}
Conjugate sextic AHO\tablenotemark[1]\tablenotetext[1]{The parameter values
in constructing the sextic AHO are taken from Table 1.} & $\ell $ & Present
work & Hill Determinant [3] & Exact Value, Eq.(9) \\
\tableline$\widehat{V}_{1}^{(4)}(r)$ & $1$ & $4.3817804600412$ & $4.381780461
$ & $4.381780459$ \\
$\widehat{V}_{1}^{(5)}(r)$ & $3$ & $7.427813527082$ & $7.427813527$ & $%
7.427813526$ \\
$\widehat{V}_{1}^{(6)}(r)$ & $5$ & $10.583005244257$ & $10.583005244$ & $%
10.583005240$%
\end{tabular}
\end{table}


\begin{references}
\bibitem{}  R. N. Chaudhuri and M. Mondal, Phys. Rev. A 40 (1989) 6080; 43
(1991) 3241.

\bibitem{}  R. K. Agrawal and V. S. Varma, Phys. Rev. A 48 (1993) 1921.

\bibitem{}  R. N. Chaudhuri and M. Mondal, Phys. Rev. A 52 (1995) 1850.

\bibitem{}  C. J. Tymczak, G. S. Japaridze, C. R. Handy and Xiao-Qian Wang,
Phys. Rev. Lett. 80 (1998) 3673.

\bibitem{}  D. P. Datta and S. Mukherjee, J. Phys. A 15 (1982) 2369.

\bibitem{}  A. Datta, J. Dey, M. Dey and P. Ghose, Phys. Lett. 106 B (1981)
505; R. N. Chaudhuri, M. Tater and M. Znojil, J. Phys. A 20 (1987) 1401.

\bibitem{}  J. Killingbeck, Phys. Lett. 65 A (1978) 87; 67 A (1978) 13.

\bibitem{}  R. P. Saxena and V. S. Varma, J. Phys. A 15 (1982) L221; G. P.
Flessas and K. P. Das, Phys. Lett. 78 A (1980) 19; A. Khare, ibid. 83 A
(1981) 237.

\bibitem{}  E. Witten, Nucl. Phys. B 185 (1981) 513; F. Cooper and B.
Freedman, Ann. Phys. (N.Y.) 146 (1983) 262; C. V. Sukumar, J. Phys. A 18
(1985) 2917.

\bibitem{}  R. Dutt, A. Khare and U. Sukhatme, Am. J. Phys. 56 (1988) 163.

\bibitem{}  V. A. Kostelecky and M. M. Nieto, Phys. Rev. Lett. 53 (1984)
2285; Phys. Rev. A 32 (1985) 1293.

\bibitem{}  M. Bernstein and L. S. Brown, Phys. Rev. Lett. 52 (1984) 1933.

\bibitem{}  T. Imbo and U. Sukhatme, Phys. Rev. Lett. 54 (1985) 2184.

\bibitem{}  A. Khare, Phys. Lett 161 B (1985) 131, R. Dutt, A. Khare and U.
Sukhatme, Phys. Lett. B 181 (1986) 295.

\bibitem{}  I. V. Dobrovolska and R. S. Tutik, Int. J. Mod. Phys. A 16
(2001) 2493; ibid. J. Phys. A 32 (1999) 563.

\bibitem{}  I. V. Dobrovolska and R. S. Tutik, [quant-ph/0611064].

\bibitem{}  S.-H. Dong, Physica Scripta 65 (2002) 289.

\bibitem{}  S. M. Ikhdair and R. Sever{\it ,} [quant-ph/0702052] submitted
to J. Molec. Spectrosc.

\bibitem{}  S.M. Ikhdair and R. Sever, J. Mol. Struct.: THEOCHEM 806 (2007)
155.

\bibitem{}  S. M. Ikhdair and R. Sever, [arXiv:quant-ph/0611065]: to appear
in Int. J. Mod. Phys. E.

\bibitem{}  S. M. Ikhdair and R. Sever{\it ,} Z. Phys. C 56 (1992) 155; C 58
(1993) 153; D 28 (1993) 1; Hadronic J. 15 (1992) 389; Int. J. Mod. Phys. A
18 (2003) 4215; A 19 (2004) 1771; A 20 (2005) 4035; A 20 (2005) 6509; A 21
(2006) 2191; A 21 (2006) 3989; A 21 (2006) 6699; [arXiv:hep-ph/0504176] (in
press) Int. J. Mod. Phys. E; S. Ikhdair, O. Mustafa and R. Sever, Tr. J.
Phys. 16 (1992) 510; 17 (1993) 474.

\bibitem{}  O. \"{O}zer and B. G\"{o}n\"{u}l, Mod. Phys. Lett. A 18 (2003)
2581.

\bibitem{}  F. Cooper, A. Khare and U. Sukhatme, Phys. Rep. 251 (1995) 267.

\bibitem{}  S. Fl\"{u}gge, Practical Quantum Mechanics 1
(Berlin-Heidelberg-New York: springer-verlag, 1971).

\bibitem{}  S. M. Ikhdair and R. Sever, Int. J. Mod. Phys. A 21 (2006) 6465;
[arXiv:quant-ph/0604073]: DOI 10.1007/s10910-006-9080-2, to appear in J.
Math. Chem.; [arXiv:quant-ph/0604078]: DOI 10.1007/s10910-006-9115-8, to
appear in J. Math. Chem.; [arXiv:quant-ph/0508009] to appear in J. Math.
Chem ; [arXiv:quant-ph/0603205] J. Mol. Struct. (THEOCHEM) (2007)
(DOI:10.1016/j.theochem.2007.01.019).
\end{references}
\end{document}